\documentclass[12pt]{article}
\usepackage{amssymb,amsmath}

\usepackage{curves}
\usepackage{epic}
\usepackage{eepic}
\usepackage{epsfig}

\newcommand{\dd}{\mbox{\rm d}}

\newcommand{\DD}{\mbox{\rm D}}

\newcommand{\p}{\partial}

\newcommand{\be}{\begin{equation}}
\newcommand{\bear}{\begin{eqnarray}}

\newcommand{\ear}{\end{eqnarray}}
\newcommand{\ee}{\end{equation}}
\newcommand{\lbl}{\label}
\newcommand{\bi}{\bibitem}
\newcommand{\ci}{\cite}

\newcommand{\vs}{\vspace}
\newcommand{\hs}{\hspace}

\begin{document}

\begin{center}

\
\vs{1cm}

\baselineskip .7cm

{\bf \Large Klein-Gordon-Wheeler-DeWitt-Schr\"odinger Equation} \\

\vs{2mm}

\baselineskip .5cm
Matej Pav\v si\v c

Jo\v zef Stefan Institute, Jamova 39, SI-1000, Ljubljana, Slovenia; 

email: matej.pavsic@ijs.si

\vs{3mm}

{\bf Abstract}
\end{center}

\baselineskip .43cm
{\small

We start from the Einstein-Hilbert action for the gravitational
field in the presence of a ``point particle'' source, and cast the
action into the corresponding
phase space form. The dynamical variables of such a system satisfy
the point particle mass shell constraint,  the Hamilton and the momentum
constraints of the canonical gravity. In the quantized theory, those
constraints become operators that annihilate a state. A state can
be represented by a wave functional $\Psi$ that simultaneously satisfies
the Klein-Gordon and the Wheeler-DeWitt-Schr\"odinger equation. The
latter equation, besides the term due to gravity, also  contains the
Schr\"odinger like term, namely the derivative of $\Psi$ with respect
to time, that occurs because of the presence of the point particle.
The particle's time coordinate, $X^0$, serves the role of time.
Next, we generalize the system to $p$-branes, and find out that for
a quantized spacetime filling brane there occurs an effective cosmological
constant, proportional to the expectation value of the brane's momentum,
a degree of freedom that has two discrete values only, a positive and
a negative one.
This mechanism could be an explanation for the small cosmological constant
that drives the accelerated expansion of the universe.

}

\baselineskip .55cm

\section{Introduction}

The meaning of time in quantum gravity is still a matter of debate
(for a recent review see\,\ci{Anderson}). A possible resolution of this
problem is to consider matter degrees of freedom from which, upon
quantization, one can obtain the derivative of the wave functional
with respect to a time variable\,\ci{Rovelli} in the Wheeler-DeWitt
equation\,\ci{WheelerDeWitt}. The idea is to introduce a reference
fluid\,\ci{Fluid}, which enables the identification of spacetime points
and the occurrence of a time variable. Instead of a fluid, one can consider
a model with one point particle only\,\ci{Rovelli}. In this letter we will
further explore and adapt that model. We start with the Einstein-Hilbert
action for gravity in the presence of a ``point particle'' source that is
in fact an extended object, like a ball, whose center of mass worldline
satisfies the equations of motion for a point particle. Then we
cast the action into the phase space form that involves the set of
Lagrange multipliers: $\alpha$, the einbein  on the particle's worldline,
$N$, the lapse and $N^i$, the shift functions that occur in the ADM
decomposition\,\ci{ADM} of the spacetime metric tensor $g_{\mu \nu}$. Variation of
the action with respect to $\alpha$, $N$ and $N^i$ gives the mass shell
constraint, the Hamilton and the momentum constraint.
In the quantized theory, such
system is described by a wave functional $\Psi[X^\mu,q_{ij}]$ that satisfies
the Klein-Gordon and the Wheeler-DeWitt-Schr\"odinger equation. The latter
equation contains, besides the usual Wheeler-DeWitt terms due
to gravity, also the term $\delta^3 ({\bf x} - {\bf X})\, i \p \Psi/\p X^0$
due to the point particle. In addition, the wave functional also satisfies
the quantum momentum constraint that contains the term
$\delta^3 ({\bf x} - {\bf X})\, i \p \Psi/\p X^i$, $i=1,2,3$.

In distinction to Rovelli, we do not introduce here an extra, the so called
``clock dynamical variable", associated with the particle. In our approach
we use the time component $x^0 = X^0(\tau)$ of the worldline parametric
equation $x^\mu = X^\mu (\tau)$, and fix the parameter $\tau$ by requiring
$X^0 (\tau) = \tau$. It turns out that the particle coordinate $X^0$
serves as evolution parameter, just like in field theories. That $X^0$,
which is not the particle dynamical degree of freedom, serves as time is
in agreement with the well known fact that time in quantum mechanics is
not a dynamical degree of freedom, but merely a parameter. According to
this line of reasoning, we do not need to worry how to find a dynamical
variable with the role of time. It comes out that $t\equiv x^0 = X^0$,
i.e., the quantity that in special and general relativity we anyway call
``time" , is indeed time, since it can serve as an evolution parameter.
This happens, if we do not consider gravity in empty space, but gravity
in the presence of a point particle for which it is no problem to
identify $X^0$ as time. 

Next, we consider the gravity in the presence of many particles, and
finally in the presence of a $p$-brane. Then, instead of one time,
we have the many fingered time $X^0 (\sigma^a)$, $a=1,2,...,p$. The
wave functional for the brane satisfies, besides the
Wheeler-DeWitt-Schr\"odinger equation, also the quantum $p$-brane
constraints that replace the Klein-Gordon equation.
We explore a special case of a spacetime filling brane, and
obtain the positive or negative cosmological constant that depends
on the sign of the brane momentum $p_0$. The latter momentum, because
of the $p$-brane constraint, has two discrete values only,
$p_0 = + \mu_B \sqrt{q}$ and $p_0 = - \mu_B \sqrt{q}$, where $\mu_B$
is the brane tension and $q$ the determinant of the 3-space metric.
The quantized theory then gives an expectation value $\langle \hat p_0 \rangle$
for the  state that is a superposition of the eigenstates with positive
and negative $p_0$. The effective cosmological constant, proportional
to $\langle \hat p_0 \rangle$, has thus a continuous range of possible values,
including the one that fits the observed accelerated expansion of the
universe. At the end we discuss the possibility that a 3-brane in a higher
dimensional bulk space is our world---a ``brane world.''

\section{The Einstein-Hilbert action with a ``point particle'' matter term
and its quantization}

Let us consider the Einstein-Hilbert action for the gravitational field
$g_{\mu \nu} (x)$, $\mu,~\nu = 0,1,2,3$, in the presence of a ``point
particle'' source, described by variables $X^\mu (\tau)$:
\be
\,\,I[X^\mu ,g_{\mu \nu} ] = m \int {\rm{d}} \tau \,
(\dot X^\mu \dot X^\nu g_{\mu \nu} )^{1/2}  
+ \,\,\kappa \int {\rm{d}}^4 x\, \sqrt{-g} \, R \,\,
\lbl{2.1}
\ee
where $\kappa \equiv {1}/(16\pi G)$. It is well known that the Einstein
equations with a point like source have no solution, because a solution in
in the vacuum around a source is the black hole with a horizon, the black
hole singularity being spacelike and cannot hence be interpreted as a
point particle worldline. However, for the sake of completeness, let
me mention that alternative views can be found in the
literature\,\ci{CastroPoint}. Leaving such intricacies aside, we can
nevertheless use the action (\ref{2.1}) as an approximation to a realistic
physical situation in which instead of a point particle we have an
extended source, described by $X^\mu (\tau, \sigma^a)$, with $X^\mu (\tau)$
being the center of mass coordinates. In particular,
if the particle is a ball, then the parameters are $\sigma^a =(R,\theta,\phi)$,
$0<R<R_0$, $0 <\theta <\pi$, $0<\phi<2 \pi$, where $R_0$ is greater than the
Schwarzschild radius.

Let us now consider the ADM split of spacetime, 
$M_{1,3}=\mathbb{R} \times \mathbb{R}^{0,3}$. Then the 4D metric can be
decomposed as
\be
   g_{\mu \nu} = \begin{pmatrix}  N^2-N^i N_i , & -N_i& \\
                    -N_j ,& - q_{ij} \\
                    \end{pmatrix}
\lbl{2.2a}
\ee 
where $N=\sqrt{1/g^{00}}$ and $N_i = - g_{0 i}$, $i=1,2,3$, are the laps and
shift functions. The inverse metric is
\be
   g^{\mu \nu} = \begin{pmatrix}  {1}/{N^2} , & -{N^i}/{N^2}& \\
                    -{N^j}/{N^2} ,& {N^i N^j}/{N^2}-q^{ij}\\
                    \end{pmatrix}
\lbl{2.2b}
\ee
Here $q^{ij}$ is the inverse of $q_{ij}$ and $N^i=q^{ij} N_j$.

The gravitational part of the action (\ref{2.1}) can be cast, by using
Ref.\,\ci{WheelerDeWitt}, into the phase space form\,\ci{GravAction}:
\be
I_G [q_{i j} ,\pi^{i j} ,N, N_i ] 
= \int {d t \,d^3 x\,
\left[ {\pi^{i j} \,\dot q_{i j} \, 
- \,N {\cal H} (q_{i j} ,\pi^{ij} )\, 
- \,N_i  {\cal H}^i  (q_{ij } ,\pi^{ij} )} \right]} ,
\lbl{2.3}
\ee
where
\bear
  && {\cal H} = - \frac{1}{\kappa}\, G_{ij \, k \ell} \pi^{ij} \pi^{k \ell}
  + \kappa \sqrt{q} R^{(3)} \lbl{2.4}\\
  &&{\cal H}^i = - 2 \DD_j \pi^{ij} \lbl{2.5}
\ear
and
\be
   G_{ij\, k \ell} = \frac{1}{2 \sqrt{q}}\, (q_{ik}q_{j \ell} + q_{i \ell}q_{jk}
   -q_{ij} q_{k \ell})
\lbl{2.6}
\ee
is the Wheeler-DeWitt metric.   
If we vary the gravitational action with respect to $N$, $N_i$, we obtain the
constraints
\bear
    &&{\cal H} = 0 \lbl{2.7} \\
    &&{\cal H}^i = 0  \lbl{2.8}
\ear
Variation with respect to $\pi^{ij}$ gives the relation
\be
     \pi^{ij} = \kappa \sqrt{q} (K^{ij} - K q^{ij})
\lbl{2.9}
\ee
where
\be
     K_{ij} = \frac{1}{2N} (\DD_i N_j + \DD_j N_i - \dot q_{ij})
\lbl{2.10}
\ee

The matter part of the action can also be cast into the phase space form:
\be
I_{\rm{m}}[X^\mu,p_\mu,\alpha] \, = \,\,\int \dd \tau \left [ {p_\mu \dot X^\mu \, 
- \,\frac{\alpha }{2}(g_{\mu \nu} p^\mu p^\nu \, - \,\,m^2 )} \right ]
\lbl{2.11}
\ee
To cast the matter part into a form comparable to the gravitational
part of the action, we insert the integration over
$\delta^4 (x - X(\tau ))d^4 x$, which gives identity. In both parts
of the action, $I_m$ and $I_G$, now stands the integration over
$\dd^4 x$. We identify $x^0 \equiv t$.

Splitting the metric according to (\ref{2.2a}), we have
\bear
   &&I_{\rm{m}}[X^\mu,p_\mu,\alpha,N,N_i,q_{ij}]  \nonumber\\
  && \hs{5mm} = \,\,\int \dd \tau 
\left( p_\mu  \dot X^\mu \, 
- \,\frac{\alpha }{2}\,\left[N^2 (p^0)^2 - q_{ij} (p^i
+ N^i p^0)\,(p^j   + N^j p^0 )  - \,m^2  \right] \right)
\lbl{2.12}
\ear
Varying the total action
\be
     I = I_{\rm G} + I_{\rm m}
\lbl{2.13}
\ee
with respect to $\alpha$, $N$ and $N^i$ we obtain the following
constraints\footnote{Since a realistic source is extended, e.g. like a ``ball",
$\int \dd \tau \delta^4 (x-X(\tau))$ should be considered as an approximation
to $\int \dd \tau \dd^3 \sigma \delta^4 (x-X(\tau,\sigma^a))$, so that, e.g.,
the constraint
${\cal H} = - \delta^3 ({\bf x}-{\bf X}) N p^0$ is an approximation to
${\cal H} = - \int \dd^3 \sigma \delta^3 ({\bf x}-{\bf X(\sigma^a})) N p^0$.
}:
\bear
  &&\delta \alpha\, : ~~~~~~~  N^2 (p^0)^2 - q_{ij} (p^i+ N^i p^0)\,
  (p^j   + N^j p^0 )  - \,m^2 = 0 \lbl{2.14} \\
  &&\delta N \, : ~~~~~{\cal H} = \int \dd \tau \, \alpha N \delta^4(x-X(\tau))
  (p^0)^2 \nonumber \\
  &&\hs{2.2cm} = - \delta^3 ({\bf x}-{\bf X}) N p^0 \lbl{2.15} \\
  &&\delta N^i\,:~~~~ {\cal H}_i = \int \dd \tau \, \alpha N \delta^4(x-X(\tau))
      q_{ij} (p^j+N^j p^0) p^0 \nonumber \\
     &&\hs{2.2cm}\, = \delta^3 ({\bf x}-{\bf X}) q_{ij} (p^j+N^j p^0) \lbl{2.16}
\ear
where ${\cal H}$ and ${\cal H}_i = q_{ij} {\cal H}^j$ are given in 
Eqs.\,(\ref{2.4}),(\ref{2.5}). Eq.\,(\ref{2.14}), of course, is nothing but the
ADM splitting of the mass shell constrain
\be
     g^{\mu \nu} p_\mu p_\nu - m^2 = 0
\lbl{2.14a}
\ee
In Eqs.\,(\ref{2.15}),(\ref{2.16}) we have performed the integration over
$\tau$, and used the equation $p^\mu =\dot X^\mu/\alpha$, that results from
varying the action (\ref{2.11}) with respect to $p^\mu$.

Let us now use the relations $p^\mu = g^{\mu \nu} p_\nu$ and
$p_\mu = g_{\mu \nu} p^\nu$ with the metrics (\ref{2.2a}),(\ref{2.2b}), and
rewrite (\ref{2.15}),(\ref{2.16}) into the form with covariant components
of momenta $p_0$, $p_i$:
\bear
    &&{\cal H} = - \delta^3 ({\bf x}-{\bf X}) \frac{1}{N} (p_0 - N^i p_i) \lbl{2.17}\\
    &&{\cal H}_i = - \delta^3 ({\bf x}-{\bf X}) p_i \lbl{2.17a}
\ear

In a quantized theory, the constraints (\ref{2.14})--(\ref{2.16}) become
operator equations acting on a state vector. In the Schr\"odinger
representation, in which $X^\mu$ and $q_{ij}(x)$ are diagonal, the momentum
operators are $\hat p_\mu = - i \p/\p X^\mu$ and $\hat \pi^{ij}
=-i \delta/\delta q^{ij}$. More precisely, momentum operators have to
satisfy the condition of hermiticity, therefore the latter definitions
are not quite correct in curved spaces, and have to be suitably modified. For instance, a possible
definition\,\ci{DeWittHermit} that
renders $\hat p_\mu$ hermitian, and also helps to resolve the factor
ordering ambiguity, is
$\hat p_\mu=-i \left [ \p_\mu+(-g)^{-1/4} \p_\mu (-g)^{1/4} \right ]$.
An alternative procedure was proposed
in Ref.\,\ci{PavsicOrder}. Analogous holds for $\hat \pi^{ij}$.

Choosing a gauge in which $N=1,~N^i=0$, we have
\be
    \left ( g^{\mu \nu} (X) \hat p_\mu \hat p_\nu - m^2 \right ) \Psi = 0
\lbl{2.18}
\ee
\be
   \hat{\cal H} \Psi = \delta^3 ({\bf x} - {\bf X}) \, i \frac{\p \Psi}{\p T}
\lbl{2.19}
\ee
\be
   \hat{\cal H}_i \Psi = \delta^3 ({\bf x} - {\bf X})\, i \frac{\p \Psi}{\p X^i}
\lbl{2.20}
\ee
A state vector is represented by $\Psi[T, X^i, q_{ij}({\bf x})]$ that
depends on the time parameter $T\equiv X^0$, the particle center of mass
coordinates\footnote{We assume that the coupled system actually describes
an extended particle whose center of mass coordinates are $X^\mu$. This system
can be envisaged to describe, e.g., the neutron that certainly is extended,
and yet only its center of mass coordinates can be considered in the wave
function. If we wish to use the above coupled system for description of
electron and other fundamental particles, one has to assume that they are
as well extended beyond their Schwarzschild radia. Otherwise those ``particles"
would be black holes. 
Since the underlying physical system whose description we have in
mind, is in fact extended, it has classical solutions.}
$X^i$, and the 3-metric $q_{ij} ({\bf x})$. In other words, $\Psi$ is a function
of $T,~X^i$, and a functional of $q_{ij} ({\bf x})$. It satisfies simultaneously
the Klein-Gordon equation (\ref{2.18}), the Wheeler-DeWitt-Schr\"odinger like
equation (\ref{2.19}), and the quantum momentum constraint (\ref{2.20})
in the presence of a point particle source. However, Eq.\,(\ref{2.19}) is
not the Schr\"odinger evolution equation; it is a constraint that has to
be satisfied at every point $\bf x$.  Since $\bf x$ runs over the 3-manifold,
we have in fact an infinite set of constraints.

Usually, for a quantum description
of gravity in the presence of matter, one does not take the matter action
in the form (\ref{2.11}). Instead, one takes\,\ci{Usual} for $I_m$ an action
for, e.g., a scalar or spinor field, and then attempts to quantize the total
action following the established procedure of quantum field theory.
Here I have pointed out that we can nevertheless start from the point
particle action (\ref{2.11}) together with the corresponding gravitational
action (\ref{2.3}). After quantization, we arrive at the Klein-Gordon equation
(\ref{2.18}) and the equations (\ref{2.19}),(\ref{2.20}) that are the
Wheeler-DeWitt equation, and the momentum constraint,  with the terms due to
the presence of point particle source. 

The presence of the $\delta$-distribution can be avoided, if we perform the
Fourier transform. The classical constraints (\ref{2.17}),(\ref{2.17a}),
with $N=1,~N^i =0$, then become
\be
    H({\bf k}) = -{\rm e}^{i {\bf k}{\bf X}} p_0
\lbl{2.21}
\ee
\be    
    H_i({\bf k}) = -{\rm e}^{i {\bf k}{\bf X}} p_i
\lbl{2.22}
\ee
where
\be
    H({\bf k}) = \int \dd^3 x\, {\rm e}^{i {\bf k}{\bf x}}\,{\cal H}~,
    ~~~~~~~~~~~~~~~
   H_i({\bf k}) = \int \dd^3 x\, {\rm e}^{i {\bf k}{\bf x}}\,{\cal H}_i
\lbl{2.22a}
\ee
and ${\bf X} \equiv X^i,~i=1,2,3$, is the particle's position at fixed
time $T$. Notice that ${\bf k}\equiv k^i$ are the Fourier partners of the
spacetime coordinates ${\bf x}$, not of the particle position ${\bf X}$.

The quantum constraint are\footnote{We are not interested here in the
issues of hermiticity and factor ordering, therefore the expressions
with $-i \delta/\delta q_{ij}$ have symbolical meaning only. In actual
calculation one has to take  suitable hermitian operators, and choose
a factor ordering.}
\be
    \int \dd^3 x \, {\rm e}^{i {\bf k}({\bf x}-{\bf X})} 
    \left ( \frac{1}{\kappa} \, G_{ij\,k \ell} \,
    \frac{\delta^2}{\delta q_{ij} \delta q_{k \ell}}
     + \kappa \, \sqrt{q} R^{(3)} \right ) \Psi
    = i \, \frac{\p \Psi}{\p T}
\lbl{2.23}
\ee
\be
   -2 \int \dd^3 x \, {\rm e}^{i {\bf k}(\bf x - \bf X)} q_{i \ell} \,
   \DD_j \left ( - i \frac{\delta}{\delta q_{j \ell}} \right ) \Psi
  = i \frac{\p \Psi}{\p X^i}
\lbl{2.24}
\ee
The above $\bf k$-dependent set of constraints (\ref{2.23}),(\ref{2.24})
replaces the set of constraints (\ref{2.19}),(\ref{2.20}). For a fixed
$\bf k$, Eq.\,(\ref{2.23}) has the form of the Schr\"odinger equation,
with the Hamilton operator that contains the functional derivatives
$-i \delta/\delta q_{ij}$. In the Hamiltonian we 
have the integration over ${\bf x}$,  just as in the Hamiltonians
of the usual field theories.

The zero mode Schr\"odinger equation, for $\bf k = 0$, is
\be
    \int \dd^3 x \, 
    \left ( \frac{1}{\kappa} \, G_{ij\,k \ell} \,
    \frac{\delta^2}{\delta q_{ij} \delta q_{k \ell}}
     + \kappa \, \sqrt{q} R^{(3)} \right ) \Psi
    = i \, \frac{\p \Psi}{\p T}
\lbl{2.25}
\ee
A solution $\Psi = \Psi_0$ of Eq.\,(\ref{2.25}) is an approximate solution
of our dynamical system. Correction terms to $\Psi_0$ come from the contributions
from the higher modes, $\bf k \neq 0$, in Eq.\,(\ref{2.23}). Bear in mind that
the momentum constraints are a consequence\,\ci{Moncrief} of the conservation
of the Hamiltonian constraint with respect to $x^0$. 

\section{Generalization to many particle and extended sources}

If instead of one, there are many particle sources, then the matter
part of the action, $I_m$, consists of the sum over single particle sources:
\be
I_{\rm{m}}[X_n^\mu,p_{n \mu},\alpha_n] \, = \sum_n \,\,\int \dd \tau_n 
\left [ {p_{n \mu} \dot X_n^\mu \, 
- \,\frac{\alpha_n }{2}(g^{\mu \nu} p_{n\mu} p_{n\nu} \, - \,\,m_n^2 )} \right ]
\lbl{3.1}
\ee
As a consequence, in the constraints (\ref{2.17}),(\ref{2.17a}), instead
of a single $\delta$-distribution, we have a sum. The quantum equations
(\ref{2.18})--(\ref{2.20}) become
\be
    \left ( g^{\mu \nu} (X_n) \hat p_{n\mu} \hat p_{n\nu} - m_n^2 \right ) \Psi = 0
\lbl{3.2}
\ee
\be
   \hat{\cal H} \Psi = \sum_n \delta^3 ({\bf x} - {\bf X}_n)
    \, i \frac{\p \Psi}{\p T_n}
\lbl{3.3}
\ee
\be
   \hat{\cal H}_i \Psi =  \sum_n \delta^3 ({\bf x} - {\bf X}_n)
   \, i \frac{\p \Psi}{\p X_n^i}
\lbl{3.4}
\ee
The Wheeler-DeWitt equation thus becomes a multi fingered time equation.
Its Fourier transform is
\be
    \int \dd^3 {\bf x} \, {\rm e}^{i {\bf k} {\bf x}}\, {\cal H} \Psi =
    i \sum_n {\rm e}^{i {\bf k} {\bf X}_n }\, \frac{\p \Psi}{\p T_n}
\lbl{3.5}
\ee
We can single out one particle, denote its time and spatial coordinates
as $T$ and ${\bf X}$, respectively, and rewrite Eq.\,(\ref{3.5}) according
to
\be
    \int \dd^3 {\bf x} \, {\rm e}^{i {\bf k} ({\bf x}-{\bf X})}\, {\cal H} \Psi =
    i \sum_{n=1}^{N-1} {\rm e}^{i {\bf k} ({\bf X}_n -{\bf X})}
    \, \frac{\p \Psi}{\p T_n} +  i\frac{\p \Psi}{\p T}
\lbl{3.6}
\ee 
One particle, in the above case the $N^{th}$ one, was singled out and chosen
as a clock that measures a time $T$. The name `particle' in the quantum
equation (\ref{3.6}) should be taken with caution. In fact we have
a system of many gravitationally interacting particles, described by
$\Psi[T_n,{\bf X}_n, q_{ij} ({\bf x})]$, $n=1,2,...,N$, and if particles are
indistinguishable, one cannot say which particle is at which position.
What we have singled out was in fact one of the parameters $T_n$, namely
$T_N \equiv T$.

Instead of a system of many particles, we can consider an extended
source, for instance, a $p$-brane. Then we
have\,\ci{PavsicPhaseBrane}--\ci{BarutPavsicBrane}:
\be
    I_m [X^\mu, p_\mu, \alpha, \alpha^a] = \int \dd \tau \, \dd^p \sigma
    \left [ p_\mu \dot X^\mu -\frac{\alpha}{2 \mu_B \sqrt{|\bar f|}}
     (g^{\mu \nu} p_\mu p_\nu + \mu_B^2 \bar f) 
     - \alpha^a \p_a X^\mu p_\mu \right ]
\lbl{3.7}
\ee
Here $\mu_B$ is the brane tension, $\tau,~\sigma^a$, $a=1,2,...,p$, the brane
time like and space like parameters, $\alpha,~\alpha^a$, Lagrange multipliers,
$i,~j=1,2,...,D-1$, the spatial indices of the $D$-dimensional
spacetime in which the brane is embedded, and
$\bar f \equiv {\rm det}\, \bar f_{ab}$ the determinant of the induced metric
$\bar f_{ab}\equiv \p_a X^\mu \p_b X^\nu g_{\mu \nu}$.

If we vary the action (\ref{3.7}) with respect to $\alpha,~\alpha^a$, we obtain
the $p$-brane constraints\,\ci{BraneConstraints,PavsicPointBrane}:
\be
     g^{\mu \nu} p_\mu p_\nu + \mu_B^2 \bar f = 0~, ~~~~~~\p_a X^\mu p_\mu =0
\lbl{3.7b}
\ee
and if we vary the total action $I_g + I_m$ with respect to $N,~N^i$,
we obtain the constraints
\bear
  &&{\cal H} = -\int \dd^p \sigma \, p_0 \, \delta^{D-1} ({\bf x}-{\bf X}(\sigma))
     \lbl{3.7c}\\
 &&{\cal H}_i = -\int \dd^p \sigma \, p_i \, \delta^{D-1} ({\bf x}-{\bf X}(\sigma)) 
    \lbl{3.7d}
\ear     
     
Varying the action (\ref{3.7}) with respect to $p_\mu$, we obtain the relation
between momenta and velocities:
\be
      p_\mu = \frac{\mu_B \dot X_\mu \sqrt{-\bar f}}{\alpha}
\lbl{3.7e}
\ee
Squaring the latter equation and combining it with Eq.\, (\ref{3.7b}),
we obtain $\alpha^2 = \dot X^\mu \dot X^\nu g_{\mu \nu}$.

As in the case of a point particle,  there are difficulties with classical
equations of motion of the branes coupled to the gravitational field in all
cases except with the appropriate codimension\,\ci{Geroch}. But again, instead of
infinitely thin branes, we can consider thick branes, and inteprete the
distribution $\delta^{D-1} ({\bf x}-{\bf X}(\sigma))$ as an approximation
of the corresponding distribution for the thick brane. 

In the quantized theory, we replace\footnote{See footnote 1 and text after
Eq.\, (\ref{2.17a})}
$p_\mu (\sigma) \rightarrow -i \delta/\delta X^\mu (\sigma)$.
Instead of Eqs.\,(\ref{3.2})--(\ref{3.4}), we have
\be
    \left ( - g^{\mu \nu} \frac{\delta^2}{\delta X^\mu (\sigma)
     \delta X^\nu (\sigma)} + \mu_B^2 \bar f \right ) \Psi = 0 , ~~~~~
     \p_a X^\mu \frac{\delta \Psi}{\delta X^\mu (\sigma)} = 0
\lbl{3.8}
\ee     
\be
 \hat {\cal H} \Psi= i \int \dd^p \sigma \,\delta^{D-1}({\bf x}-{\bf X}(\sigma))
  \, \frac{\delta \Psi}{\delta T(\sigma)}
\lbl{3.9}
\ee
\be
   \hat {\cal H}_i \Psi= i \int \dd^p \sigma \,
    \delta^{D-1}({\bf x}-{\bf X}(\sigma))
   \, \frac{\delta \Psi}{\delta X^i (\sigma)}
\lbl{3.10}
\ee
where $T(\sigma) \equiv X^0 (\sigma)$, ${\bf X} (\sigma) \equiv X^i (\sigma)$,
and $\sigma \equiv \sigma^a$, $a=1,2,...,p$. In general, $\Psi =
\Psi[X^\mu (\sigma),q_{ij} ({\bf x})]$  $\equiv 
\Psi[T(\sigma), {\bf X} (\sigma),q_{ij} ({\bf x})]$
 Since now we have a spacetime of arbitrary dimension $D$, the definitions
 (\ref{2.4}),(\ref{2.5}) of ${\cal H}$ and 
${\cal H}_i$ have to be modified accordingly: $R^{(3)}$ should be replaced
by $R^{(D-1)}$, and the Wheeler-DeWitt metric is now
$G_{ij \, k \ell}=(1/(2 \sqrt{q}))[q_{ik}q_{j \ell} + q_{i \ell} q_{jk}-
(2/(D-2)) q_{ij} q_{k \ell}]$.

Of particular interest are the following special cases:

(\ i) {\it The spacetime filling brane}. Then $p=D-1$, and $i=a=1,2,...,D-1$.
One can choose  a
parametrization of $\sigma^a$ such that $X^i (\sigma) = {\delta^i}_a \sigma^a$.
Then we have  $\p_a X^i (\sigma) = {\delta^i}_a$.
  The second constraint (\ref{3.8}) then reads
$\delta \Psi/\delta X^i = 0$. This means that
$\Psi = \Psi [T(\sigma),q_{ij} ({\bf x})]$, i.e., it does not depend
on spatial functions $X^i (\sigma)$. Therefore, the first constraint (\ref{3.8})
retains the $T$-derivatives only:
\be
    \left ( - g^{00} \frac{\delta^2}{\delta T (\sigma)
     \delta T (\sigma)} + \mu_B^2 \bar f \right ) \Psi = 0 
\lbl{3.8a}
\ee

 If we now assume $\p_a T (\sigma)=0$,
the functional derivative   can be replaced by the partial derivative according
to the relation  $\delta/\delta T(\sigma) \rightarrow \sqrt{-\bar f}\,\p/\p T$.
Then, instead of (\ref{3.8a}), we have
\be
    - \frac{1}{N}\frac{\p}{\p T}\left (\frac{1}{N} \frac{\p \Psi}{\p T}
    \right ) - \mu_B^2 \Psi = 0
\lbl{3.11}
\ee
The factor ordering has been chosen in order to achieve covariance in
the one dimensional space comprised of $T$. 
The constraints (\ref{3.9}),(\ref{3.10}) become
\be
     \hat {\cal H} \Psi = \sqrt{q}\,\,
 i \frac{\p \Psi}{\p T} , ~~~~~~~~~~~
     \hat {\cal H}_i \Psi= 0
\lbl{3.12}
\ee
where we have taken into account the relation
$\bar f_{ab}\equiv \p_a X^\mu \p_b X^\nu g_{\mu \nu}=g_{ab}=- q_{ab}= -q_{ij}$,
and $\bar f = -q \equiv -{\rm det}\, q_{ij}$.

Eq.\,(\ref{3.11}), in which we take $N=1$, implies that a general solution 
$\Psi[T,q_{ij}({\bf x})]$ is
a superposition of particular solutions $\Psi_+={\rm e}^{+i \mu_b T} 
\psi[q_{ij}({\bf x})]$ and
$\Psi_-={\rm e}^{-i \mu_b T} \psi[q_{ij}({\bf x})]$ that are eigenfunctions of
the operator $\hat p_0/\sqrt{q} =- i \p/\p T$ with eigenvalues $\pm \mu_B$.
For such particular solutions, the quantum Hamilton constraint equation becomes
\be
   \hat {\cal H} \Psi_\pm = \mp \sqrt{q}\, \mu_b \Psi_\pm
\lbl{3.13}
\ee
The expectation value of the operator $\hat p_0/\sqrt{q}$ in a superposition
state $\Psi = \alpha \Psi_+ + \beta \Psi_-$
is $\langle \hat p_0/\sqrt{q} \rangle =
(|\alpha|^2 - |\beta|^2) \mu_B$, where $|\alpha|^2 + |\beta|^2 =1$.

That there must be plus or minus sign in Eq.\,(\ref{3.13}),
can be seen already at the classical
level. For a spacetime filling brane, the Hamilton constraint (\ref{3.7c})
becomes ${\cal H} = - p_0$.  From Eq.\,(\ref{3.7e}) we have
$p_0 = \mu_B \dot X^\mu \sqrt{q}/\alpha$, where $\alpha 
=\sqrt{\dot X^0 \dot X^0 g_{00}}$ $= |\dot X^0| \sqrt{g_{00}} = |\dot X^0|$
is taken to be a positive quantity.
In the last step we have used $g_{00}= N^2 - N^i N_j$ and  set $N=1$,
$N^i=0$. Thus we obtain that $p_0 =  \mu_B \sqrt{q}\, \dot X^0/|\dot X^0|
= \pm \mu_B \sqrt{q}$, depending on whether $\dot X^0$ is positive or negative.
In other words, the sign of $p_0$ depends on whether the spacetime filling
brane moves forward or backwards in time. Despite that the momentum
$p_0$ of a spacetime filling brane, because of the constraint
(\ref{3.7b}), which now
reads $p_0^2 = \mu_b^2 q$, is not a continuous dynamical degree of freedom, there
still remains a freedom for $p_0$ to be either positive or negative,
more precisely, to be $p_0= \mu_b \, \sqrt{q}$ or $p_0=-\mu_B\, \sqrt{q}$.

It is illustrative to look at the situation from another angle. The
$p$-brane phase space action (\ref{3.7}) is equivalent to the minimal
surface action
\be
I_m[X^\mu (\xi)] = \mu_B \int \dd^{p+1} \xi \,
\sqrt{-{\rm det} \, \p_A X^\mu \p_B X^\nu g_{\mu \nu}}
\lbl{3.7a}
\ee
where $\xi^A = (\tau, \sigma^a)$. Performing the
ADM split on the brane's world manifold, this can be written
as\,\ci{PavsicPointBrane}
\be
   I_m [X^0,X^i] = \mu_B \int \dd \tau \, \dd^p \sigma 
   \sqrt{\dot X^\mu \dot X^\nu g_{\mu \nu}} 
   \sqrt{-{\bar f}}
\lbl{3.14}
\ee
For a spacetime filling brane we have $X^i = {\delta^i}_a \sigma^a$, $D=p+1$, and
the latter action becomes
\bear
   I_m [X^0]&=& \mu_B \int \dd \tau \, \dd^p \sigma\, 
   \sqrt{\dot X^0 \dot X^0 g_{00}} \, \sqrt{q}
   =\mu_B \int \dd X^0 \, \dd^p {\bf X}\, \frac{|\dot X^0|}{\dot X^0}
    \, N \sqrt{q} \nonumber \\
     &=& \pm \mu_B \int \dd x^0 \, \dd^p {\bf x}\, N\sqrt{q}
\lbl{3.15}
\ear
where we have used $g_{00} =N^2,~ N^i=0,~ 
\bar f = -q$, and identified $X^\mu$ with $x^\mu$.
The variation of the latter action with respect to $N$ gives
\be
     \frac{\delta I_m}{\delta N} =  \pm\sqrt{q}\, \mu_B
\lbl{3.16}
\ee     
The function $X^0(\tau)$, where $\tau$ is a monotonically increasing parameter,
has no physical meaning; it depends on choice of
coordinates. Therefore, the derivative $\dot X^0$ has no physical meaning as
well.  However, there exist two possibilities. One possibility is that
$X^0(\tau)$ increases with $\tau$.
Another possibility is that $X^0(\tau)$ decreases with $\tau$ We assume that
these two different possibilities
correspond to physically different situations, because they lead, respectively,
to the positive and negative cosmological constant. They provide an
explanation for the positive or negative sign in Eq.\,(\ref{3.13}).

We can look at the situation even more directly. Since in the case of a spacetime
filling brane its world volume fils the embedding spacetime, we can choose
coordinates in the action (\ref{3.7a})  so that
$X^\mu (\xi^A) = {\delta^\mu}_A \xi^A = \xi^\mu$. Bear in mind that now
$\mu=0,1,2,...,D-1$ and $A=0,1,2,...,p=D-1$. By such choice of coordinates,
we obtain $I_m = \mu_B \, \int \dd^4 x \, \sqrt{-g}$. But we may as well
choose $X^0 (\xi^A) = - \tau$, where $\tau \equiv \xi^0$, which means that
$X^0$ increases in the opposite direction than $\tau$ does, and so,
figuratively speaking, our brane ``moves backward in time". Then we obtain
$I_m = - \mu_B \int \dd^4 x \, \sqrt{-g}$. This corresponds to the 
term with the cosmological constant $\Lambda = \pm 16 \pi G\, \mu_B
=16 \pi G \, p_0/\sqrt{q}$.
The spacetime filling brane is thus responsible for the cosmological constant,
which can be positive or negative. In the quantized theory, a generic state
is a superposition of those two possibilities, 
$\Psi= \alpha \Psi_+ + \beta \, \Psi_-$,
where the eigenstates $\Psi_{\pm}$ simultaneously satisfy Eq.\,(\ref{3.11})
and (\ref{3.12}). We have thus verified that, in the case of the spacetime
filling brane the system of equations  (\ref{3.8})--(\ref{3.10}) has
a consistent solution. For the expectation value
of the cosmological constant in the superposition state 
state $\Psi$ we obtain
\be
\langle \hat \Lambda \rangle =
16 \pi G \langle \frac{\hat p_0} {\sqrt{q}}\rangle = (|\alpha|^2 - |\beta|^2)
16 \pi G \mu_B
\lbl{3.17}
\ee
It can be any value between $16 \pi G \mu_B$ and $-16 \pi G \mu_B$,
including zero or a small value that fits the accelerated expansion of the
universe.
 That a spacetime filling brane gives the cosmological
constant was considered by Bandos\,\ci{Bandos}, but he took into
account one sign only.

   (ii) {\it The brane is a brane world.}

Another possibility is that a 3-brane, embedded in a higher dimensional
spacetime (bulk), is our observable world
(``brane world")\,\ci{BraneWorld}.
Then everything that directly counts for us as observers are points on
the brane. It does not matter that points outside the brane cannot be
identified, and that Eqs.\,(\ref{3.9}),(\ref{3.10}) read 
${\cal H}\Psi=0$, ${\cal H}_i \Psi =0$, which implies that the wave functional 
is "timeless", with no evolution. What matters is that
the wave functional on the brane, i.e., at ${\bf x} = {\bf X}(\sigma)$,
has evolution due to the term with $\delta \Psi/\delta T(\sigma)$ on the
right hand side of Eq.\,(\ref{3.9}). However, strictly speaking,
Eq.\,(\ref{3.9}) is not a true evolution (Schr\"odinger)
equation; it is a set of constraints, valid at any point ${\bf x}$, that can
be on the brane or in the bulk. If we perform the Fourier transform,
then we obtain the brane analogue of Eq.\,(\ref{2.23}), and
the zero mode  equation has the form of the Schr\"odinger equation,
with the Hamilton operator $H|_{{\bf k}=0}=\int \dd^3 {\bf x}\, {\cal H}$.
Because of
the integration over ${\bf x}$, the quantity $H|_{{\bf k}=0}$ has the correct
form of a field theoretic Hamiltonian. However, such zero mode Schr\"odinger
equation does not provide a complete description of the system, for which
also all higher modes with ${\bf k}\neq 0$ are necessary.

\section {Discussion and conclusion}

We have considered a ``point particle'' coupled to the gravitational field.
The classical constraints become after quantizations a system of
equations that comprises the Klein-Gordon, Wheeler-DeWitt and
Schr\"odinger equation. Then we generalize the theory to $p$-branes, in which
case the Klein-Gordon equation is replaced by the $p$-brane quantum
constraints. In our approach we start from a classical theory in which
the point particle or the brane coordinates $X^\mu$ and the spatial
metric $q_{ij}$ are on the same footing in the sense that they are
the quantities that describe the system. In the quantized theory, the
system is described by a wave functional $\Psi[X^\mu,q_{ij}]$ that satisfies
the system of equations (\ref{3.8}--(\ref{3.10}). In the case of a point
particle, the latter system becomes (\ref{2.18})--(\ref{2.20}).
A benefit of such approach is that there is no problem of time.
The matter coordinate $X^0 \equiv T$ is time. Moreover, the
Wheeler-DeWitt equation has the part $i \p \Psi/\p T$, just as
the Schr\"odinger equation.

The wave function(al) $\Psi[X^\mu, q_{ij}]$, satisfying the Klein-Gordon
equation, is a generalization of the Klein-Gordon field that depends
on $X^\mu$ only. In quantum field theory,
the Klein-Gordon field, after second quantization, becomes an
operator field that,
roughly speaking, creates and annihilates particles at spacetime points $X^\mu$.
Analogously, we can envisage, that the function(al) $\Psi[X^\mu, q_{ij}]$
should also be considered as a field that can be (secondly) quantized and
promoted to an operator that creates or annihilates a particle
(in general, a $p$-brane) at  $X^\mu$, together with the gravitational
field $q_{ij}$. An action functional for $\Psi[X^\mu, q_{ij}]$ that leads
to Eqs.\,(\ref{2.18})--(\ref{2.20}) should be found and its quantization
carried out, together with the calculation of the corresponding vacuum
energy density due to
the quantum field $\hat \psi [X^\mu, q_{ij}]$. It has to be investigated
anew, how within such generalized theory
vacuum energy influences the gravitational field and what is
its effect on the cosmological constant.

Instead of the point particle action (\ref{2.11}) that leads to the
Klein-Gordon equation, we could take the spinning particle
action\,\ci{Townsend} that leads to the Dirac equation. Then, in the
system (\ref{2.18})--(\ref{2.20}) we would have the Dirac instead of
the Klein-Gordon equation, and $\Psi[X^\mu, q_{ij}]$ would be a
generalization of the Dirac field.

We have thus a vision that the quantum field theory of a scalar or spinor
field in the presence of a gravitational field could  be
formulated differently from what we have been accustomed so far.
Usually, we have an $x^\mu$-dependent field, e.g., a scalar
field $\varphi (x)$ or a spinor field  $\psi (x)$, that is a ``source"
of the gravitational field $g_{\mu \nu} (x)$, decomposed, according to
ADM, into $N(x)$, $N^i (x)$ and $q_{ij}(x)$. The action is a functional
of those fields, e.g., $I[\varphi(x),N,N^i, q_{ij} (x)]$ or
$I[\psi(x),N,N^i, q_{ij} (x)]$, and in the quantized theory we have a
wave functional $\Psi[\varphi(x), q_{ij} (x)]$ or $\Psi[\psi(x), q_{ij} (x)]$.
In this paper, we investigated an alternative approach, in which the classical action was $I[X^\mu (\tau),N,N^i,q_{ij} (x)]$,
and, after quantizing it, we arrived at the wave functional
$\Psi[X^\mu,q_{ij}({\bf x})]$, i.e., a generalized field that did not depend on the particle's position $X^\mu$ in spacetime only,
but also on the dynamical variables of gravity, $q_{ij} ({\bf x})$. Quantum
field theory of the generalized field $\Psi[X^\mu,q_{ij}({\bf x})]$ is an
alternative to the usual quantum field theoretic approaches to gravity
coupled to matter. Since the usual approaches have not yet led us to
a consistent theory of quantum gravity, it is worth to investigate
what will bring the new approach, conceived in this Letter.

We also considered the case in which,
instead of a point particle, $X^\mu (\tau)$, we have a brane
$X^\mu (\tau,\sigma^a)$. In a particular case of
a spacetime filling brane, we obtained positive or negative
cosmological constant, as a consequence of the fact that the brane's
momentum $p_0$ has two discrete values only, namely
$\pm \mu_B \sqrt{q}$.
In the quantized theory, a state is a superposition of those two
possibilities, and the expectation value of the operator $\hat p_0$
is proportional to the effective cosmological constant that can be
small or can vanish. The spacetime filling brane could thus be an
explanation for a small cosmological constant driving the accelerated
expansion of the universe. Finally, there is a possibility that our
world is a 3-brane moving in a higher dimensional bulk space. The wave
functional, describing such brane, satisfies the Wheeler-DeWitt equation
with a Schr\"odinger like term $i \delta \Psi/\delta T(\sigma)$ that governs
the evolution on the brane, whereas there is no evolution in the bulk.

\vs{4mm}

\centerline{\bf Acknowledgment}

This work has been supported by the Slovenian Research Agency.

\end{document}